\documentclass{appolb}
\usepackage{epsfig}

\begin{document}
\title{Review of Linac-Ring Type Collider Proposals}
\author{A.N. Akay
\address{TOBB University of Economics and Technology, Physics Section, Ankara,Turkey}\and
 H. Karadeniz
\address{Turkish Atomic Energy Authority, SANAEM, Ankara, Turkey} \and
 S. Sultansoy
\address{TOBB University of Economics and Technology, Physics Section, Ankara, Turkey;\\
Azerbaijan Academy of Science, Institute of Physics, Baku, Azerbaijan}}

\maketitle

\begin{abstract}

There are three possibly types of particle colliders schemes: familiar (well known) ring-ring colliders, less familiar however sufficiently advanced linear colliders and less familiar and less advanced linac-ring type colliders. The aim of this paper is two-fold: to present possibly complete list of papers on linac-ring type collider proposals and to emphasize the role of linac-ring type machines for future HEP research.

\end{abstract}

\section{Introduction}
Today, experimental high energy physics deal with three kind of
devices, namely, fixed target experiments (using both accelerator
and cosmic rays), collider experiments and others (including
underground detectors and so on).

Collider experiments could be classified in two manners: keeping
in mind accelerator types or colliding beams. Concerning the first
classification there are three possible types, namely, ring-ring,
linac-linac and linac-ring colliders. Second classification
includes three types, too: hadron, lepton and lepton-hadron
colliders. With regard to energy frontiers ring-ring corresponds
to hadron collisions, linac-linac corresponds to lepton collisions
and linac-ring corresponds to lepton-hadron collisions (see Table
1 and Figure 1). Ring-ring colliders are the most advanced ones
from accelerator technology viewpoint and widespread around the
(developed) world. Linear (linac-linac) colliders are less
familiar, however, a lot of experience is handled through SLC
operation and ILC/CLIC related workout.

\begin{table}[h]
\caption{Energy Frontiers: Past and Future \cite{Sultansoy 2004}}
\begin{center}
\begin{tabular}{|l|c|c|c|}
\hline Colliders & Hadron & Lepton &
Lepton-Hadron \\ \hline 1990's & Tevatron & SLC/LEP & HERA \\
\hline $\sqrt{s}\ $(TeV) & 2 & 0.1/0.2 & 0.3 \\ \hline L
($10^{31}$ $cm^{-2}s^{-1}$) & 1 & 0.1/1 & 1 \\ \hline 2010's & LHC
& "NLC" & "NLC"-LHC \\ \hline $\sqrt{s}\ $(TeV) & 14 & 0.5 & 3.7
\\ \hline L ($10^{31}$
$cm^{-2}s^{-1}$) & $10^{3}$ & $10^{3}$ & $1\div10$ \\ \hline
2020's & VLHC & CLIC & "CLIC"-VLHC \\ \hline $\sqrt{s}\ $(TeV) &
200 & 3 & 34 \\ \hline L ($10^{31}$ $cm^{-2}s^{-1}$) & $10^{3}$ &
$10^{3}$ & $10\div100$ \\ \hline
\end{tabular}
\end{center}
\end{table}

Forty years ago John Rees proposed to collide 20 GeV SLAC electron
beam with 3 GeV stored positrons \cite{Rees 1969} to handle 15.5
GeV center-of-mass energy electron-positron collisions with a
luminosity of $5\times10^{29}$ $cm^{-2}s^{-1}$. Two years later
this proposal reconsidered in \cite{Csonka 1971} keeping in mind 2
GeV stored electrons (or positrons) which corresponds to 12.6 GeV
center-of-mass energy with a luminosity of $2.4\times10^{29}$
$cm^{-2}s^{-1}$. Both proposals were considered as possible
upgrades of SLAC accelerator \cite{Miller 1971}. During following
fifteen years only one paper is published on the subject
\cite{Csonka 1981}. The reason was choosing a linear collider
option for SLAC upgrade: the Stanford Linear Collider (SLC)
construction began in 1983 and was completed in 1989. In 1979
linac-ring scheme was considered in short as an alternative option
for SSC based ring-ring type 140 GeV + 20 TeV electron-proton
collider \cite{Weber 1979} (see also \cite{Berley 1982}).

The idea was reborn in the mid of 1980's in order to combine
linear electron-positron and ring type proton colliders had to
realize additional TeV scale lepton-hadron collider option.
Namely, it was proposed to construct VLEPP tangentially to UNK
\cite{Alekhin 1987}. This scheme will provide an opportunity to
handle TeV scale $\gamma$p colliders, too \cite{Alekhin 1991}.
This line was go on by THERA, EIC/EPIC and QCD-E/LHeC projects
(for references see corresponding sections below). An important
stage in this direction was provided by the International Workshop
held in Ankara in 1997 \cite{Ankara 1997}. There are a number of
reviews on the subject \cite{Sultanov 1989, Wiik 1993, Brinkmann
1997, Sultansoy 1998, Sultansoy 1999, Sultansoy 2004}.

\begin{figure}
\begin{center}
\includegraphics*[width=12.0cm]{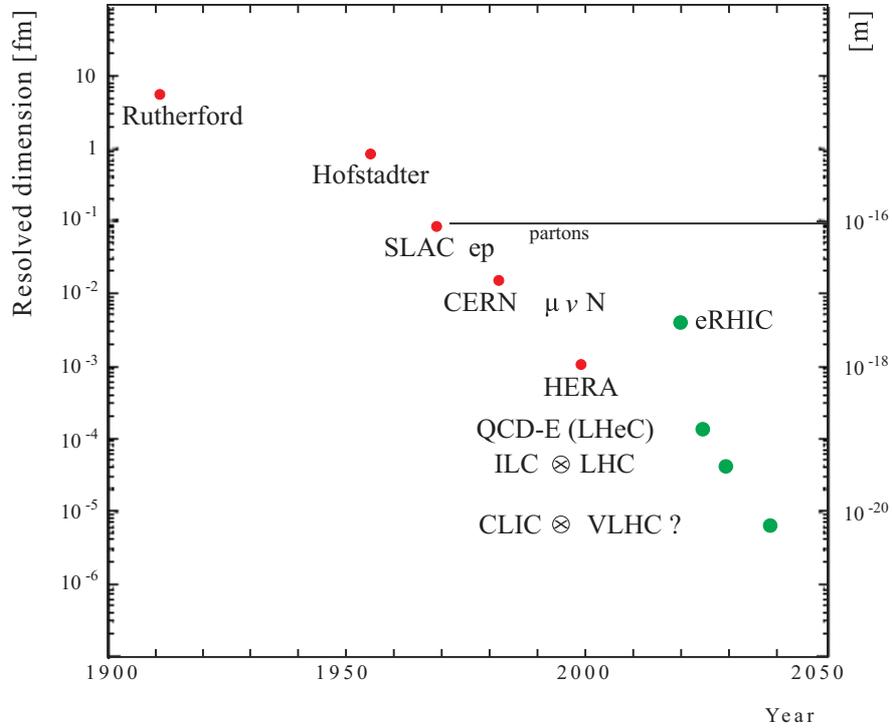}
\caption{The development of the resolution power of the
experiments exploring the inner structure of matter over time from
Rutherford experiment to $CLIC\bigotimes$VLHC.}
\end{center}
\end{figure}

Another line deals with particle factories (see Figure 2): in 1988
Grosse-Wiesmann proposed linac-ring type B-factory \cite{Wiesmann
1989, Wiesmann 1990, Cline 1990, Wiesmann 1991}. In 1993
linac-ring type charm-tau factory had proposed as the regional
project for Turkey and abroad \cite{Sultansoy 1993}. The last
stage of this line is represented by Super Charm Factory as the
part of the Turkic Accelerator Complex (TAC) Project \cite{TAC
Web}.

\begin{figure}
\begin{center}
\includegraphics*[width=14.0cm]{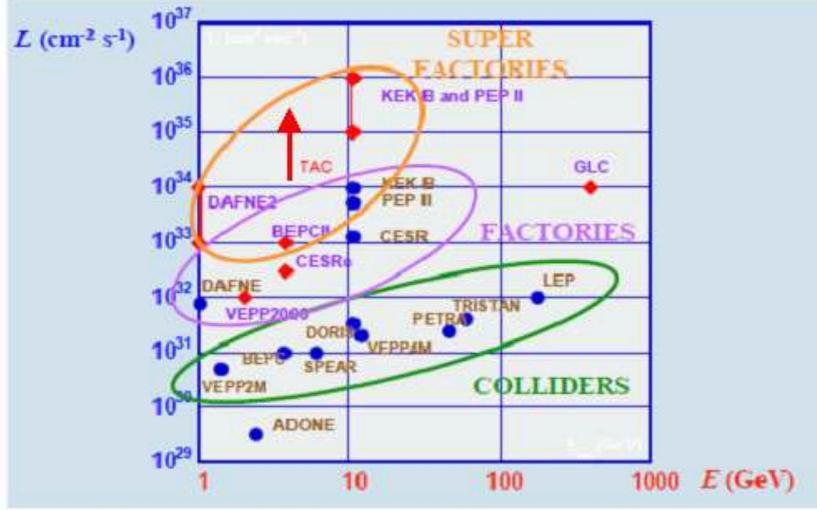}
\caption{Past, present and future $e^{+}e^{-}$ colliders.}
\end{center}
\end{figure}

The content of the review is following. In section 2, main
parameters of linac-ring type lepton-hadron collider proposals are
considered, namely, UNK+VLEPP, THERA, eRHIC, EIC, QCD Explorer
(LHeC linac-ring option) and Energy Frontier. Photon-Hadron
colliders would be constructed on the base of these colliders are
considered in section 3. Section 4 is devoted to linac-ring type
particle factory proposals. Finally, in section 5 some concluding
remarks and recommendations are presented.

\section{Lepton-Hadron Colliders}

There are a number of reasons \cite{Tigner 1991, Wiik 1993}
favoring a superconducting linear collider (such as TESLA) as a
source of e-beam for linac-ring type colliders. First of all,
spacing between bunches in warm linacs, which is of the order of
ns, doesn't match with the bunch spacing in the HERA, TEVATRON and
LHC. Also the pulse length is much shorter than the ring
circumference. In the case of TESLA, which use standing wave
cavities, one can use both shoulders of linac in order to double
electron beam energy, whereas in the case of conventional linear
colliders one can use only half of the machine, because the
traveling wave structures can accelerate only in one direction.

The most transparent expression for the luminosity of linac-ring
type ep colliders is \cite{Wiik 1993}:

\begin{equation}
\L_{ep}=\frac{1}{4\pi }\frac{P_{e}}{E_{e}}\frac{N_{p}}
{\varepsilon _{p}^{N}}%
\frac{\gamma _{p}}{\beta _{p}^{\ast }}
 \label{1}
\end{equation}

for round, transversely matched beams. The lower limit on $\beta
_{p}^{\ast}$, which is given by proton bunch length, can be
overcome by applying a "dynamic" focusing scheme \cite{Brinkmann
1995}, where the proton bunch waist travel with electron bunch
during collision. In this scheme $\beta _{p}^{\ast}$ is limited,
in principle, by the electron bunch length, which is two orders
magnitude smaller. More conservatively, an upgrade of the
luminosity by a factor 3-4 may be possible.

\subsection{UNK+VLEPP (IHEP, Protvino)}

In 1980's there were two energy frontier collider projects in the
former USSR, namely, $\sqrt{s}\ = 6$ TeV proton-proton collider
UNK and $\sqrt{s}\ = 2$ TeV linear electron-positron collider
VLEPP. The construction of the first one was started at IHEP
(Protvino, Moscow region), and the second one was planned at BINP
(Novosibirsk). In mid 1980's the construction of VLEPP tangential
to UNK was proposed in order to provide additional opportunity to
handle energy frontier ep \cite{Alekhin 1987} and $\gamma$p \cite
{Alekhin 1991} colliders.

Luminosity estimations were given in \cite{Sultanov 1989}. Brief
resume is followed. Two versions of placement of VLEPP regarding
to UNK are possible, namely symmetric (Figure 3a) and asymmetric
(Figure 3b). Two options of ep and $\gamma$p collisions were
considered for the UNK+VLEPP: on extracted proton beam (Figure 4a)
or in proton ring (Figure 4b). It was shown that L = $10^{30}$
$cm^{-2}s^{-1}$ and L = $6\times$$10^{30}$ $cm^{-2}s^{-1}$ could
be achievable for first and second option, respectively.

\begin{figure}
\includegraphics*[width=12.0cm]{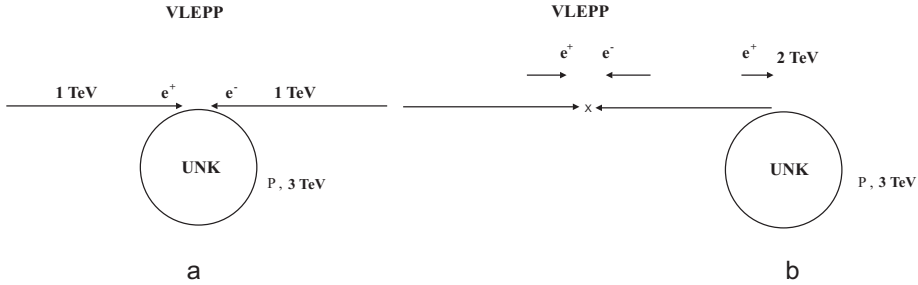}
\caption{a) Symmetric version of ep ($\gamma$p) collider; b)
Asymmetric version \cite{Sultanov 1989}.}
\end{figure}

\begin{figure}
\includegraphics*[width=12.0cm]{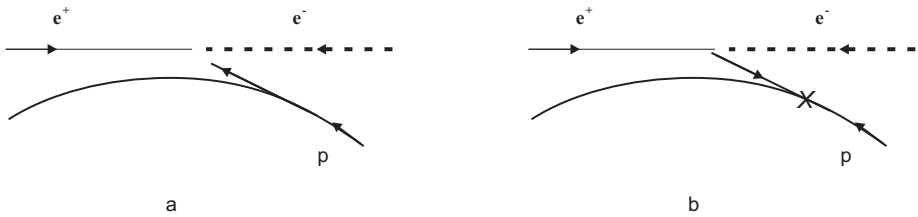}
\caption{a) ep ($\gamma$p) collisions on extracted p beam; b) ep
($\gamma$p) collisions in the proton ring \cite{Sultanov 1989}.}
\end{figure}

Note that this consideration was the main scientific reason for
moving of VLEPP from Novosibirsk to Protvino. Unfortunately, in
final design VLEPP placement was chosen to cross the UNK ring,
instead of tangential. Obviously, this choice closed ep and
$\gamma$p options (clear indication of collapse of Eastern Block).

The status of ep collider proposals in the end of 1990's is
presented in Table 2 (the consideration of the LHC+CLIC and
SSC+LSC proposals was initiated by the A. Salam).

\begin{table}[h]
\caption{Planned and possible ep colliders (as for 1989
\cite{Sultanov 1989})}
\begin{center}
\begin{tabular}{llllllll}
\hline Machine & $\sqrt{s}$ & $E_{e}$ & $E_{p}$ & $n_{e}$ &
$n_{p}$ & Coll. rate & L,$10^{30}$ \\
  & TeV & TeV & TeV & $10^{10}$ & $10^{10}$ & f(Hz) & $cm^{-2}s^{-1}$ \\
  \\
  Standart ep machines \\
  HERA & 0.3 & 0.03 & 0.82 & 3.48 & 10 & $10^{7}$ & 15 \\
  LHC+LEPI & 1.3 & 0.05 & 8 & 8.2 & 30 & $5\times10^{6}$ & 200 \\
  LHC+LEPII & 1.8 & 0.1 & 8 & 8.2 & 30 & $5\times10^{6}$ & 10 \\
  UNK+e-ring & 0.6 & 0.03 & 3 & - & - & - & 100 \\
  SSC+e-ring & 2.8 & 0.1 & 20 & - & - & - & 100 \\
  \\
  New type ep $machines^{1}$ \\
  UNK+VLEPPI & 2.4 & 0.5 & 3 & 20 & 100 & 100 & 1+10 \\
  UNK+VLEPPII & 3.5 & 1 & 3 & 20 & 100 & 100 & 1+10 \\
  UNK+$VLEPP^{2}$ & 4.9 & 2 & 3 & 20 & 100 & 100 & 1+10 \\
  LHC+$CLIC^{2}$ & 8 & 2 & 8 & 0.5 & 100 & $6\times10^{3}$ & 10+100 \\
  LHC+$CLIC^{3}$ & 4.8 & 0.7 & 8 & 0.6 & 100 & $10^{4}$ & 10+100 \\
  SSC+$LSC^{2}$ & 28 & 10 & 20 & 0.08 & 100 & $8\times10^{4}$ & 100 \\
  LHC+$e-linac^{4}$ & 3 & 0.3 & 8 & 0.08 & 100 & $5\times10^{5}$ & $5\times10^{3}$ \\
  SSC+$e-linac^{4}$ & 8 & 0.8 & 20 & 0.05 & 100 & $5\times10^{5}$ & $10^{4}$ \\
  \\
  Electron-proton $linacs^{5}$ \\
  VLEPP+p10 & 2 & 1 & 1 & 20 & 20 & 100 & 1+10 \\
  LSC+p10 & 10 & 5 & 5 & 0.08 & 0.08 & $8\times10^{4}$ & 100 \\
\hline
\end{tabular}
\end{center}
\footnotesize
\footnote{1}Given parameters corresponds to ep collisions in the
proton ring

\footnote{2}asymmetric version (see Figure 3b)

\footnote{3}ep version of Linac-LEP Collider with $\sqrt{s} = 0.5$
$TeV$ proposed by C. Rubbia (see Ref 24)

\footnote{4}P. Grosse-Wiesmann's proposal: electron linacs
parameters are optimized for ep collisions (see Ref 16)

\footnote{5}One shoulder of the $e^{+}$$e^{-}$ linac can be used
for acceleration of protons, if 10 GeV proton ring will be added
at its beginning.
\end{table}

\subsection{THERA (DESY)}

THERA activity had been initiated since 1996 by B. Wiik and S.
Sultansoy \cite{Soltansoy 1996}. A lot of work was down in
1999-2000 during preparation of the TESLA TDR, which include THERA
\cite{THERA web} as inseparable part together with photon collider
and fixed target options. Moreover, the THERA provided the main
scientific reason for moving of TESLA from Zeuthen to Hamburg.
Results of this two years study are published in the THERA Book
\cite{THERA book}. Concerning beam energies and corresponding
luminosities 3 alternatives were analyzed (see Table 3).

\begin{table}[h]
\caption{THERA beam energies and luminosities}
\begin{center}
\begin{tabular}{|c|c|c|c|c|}
\hline Option & $E_{e}$, GeV & $E_{p}$, GeV & $\sqrt{s}$, TeV &
L, $10^{30}$ $cm^{-2}s^{-1}$ \\ \hline 1 & 250 & 1000 & 1 & 4 \\
\hline 2 & 500 & 500 & 1 & 25 \\ \hline 3 & 800 & 800 & 1.6 & 16 \\
\hline
\end{tabular}
\end{center}
\end{table}

Unfortunately, the THERA paper \cite{Abramowicz 2001} were submitted to LANL archive with
wrong information (TESLA-N Study Group instead of THERA Working Group) and were not submitted for publication to
journal. This event resulted in practical zero citation comparing
with TESLA photon collider paper \cite{Badelek 2001} with more than 160 citations.

\subsection{EIC (USA)}

The Electron-Ion Collider (EIC) is the proposed new facility to
collide high-energy electrons with nuclei and polarized
protons/light nuclei \cite{Guzey 2009, Thomas 2009} and refs
therein). Two broad classes of goals of the future EIC are
reflected in two physics working groups (WG) of the EIC
collaboration: the eA WG concentrates on exploring the (strong)
gluon fields in nuclei, and the ep WG focuses on the precision
imaging of quarks and gluons in the nucleon.

The original design of the EIC involves two concepts: eRHIC on the
base of RHIC (see Figure 5), where an additional energy recovering
linac has to added, and ELIC at Jefferson Lab (see Figure 6),
which requires a construction of a new hadron facility to be used
with the existing CEBAF. The eRHIC concept allows for larger
$\sqrt{s}$ = 60-90 GeV and smaller luminosity L $\approx$
$10^{33}$ $cm^{-2}s^{-1}$, while the ELIC concept corresponds to
smaller $\sqrt{s}$ $\leq$ 60 GeV and larger luminosity L $\approx$
$10^{35}$ $cm^{-2}s^{-1}$.

For more details see EIC, eRHIC and ELIC web pages \cite{EIC web,
eRHIC web, ELIC web}.

\begin{figure}
\begin{center}
\includegraphics*[width=10.0cm]{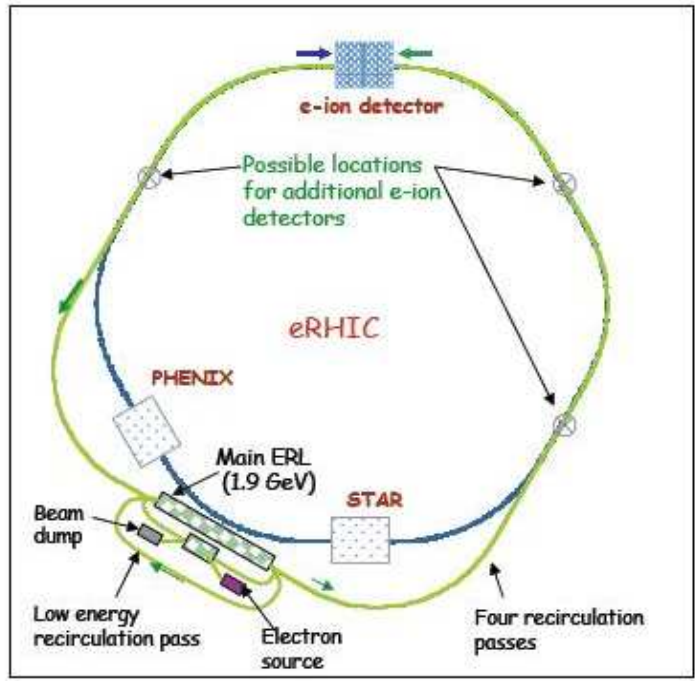}
\caption{ERL-based eRHIC design \cite{Litvinenko 2009}}
\end{center}
\end{figure}

\begin{figure}
\begin{center}
\includegraphics*[width=10.0cm]{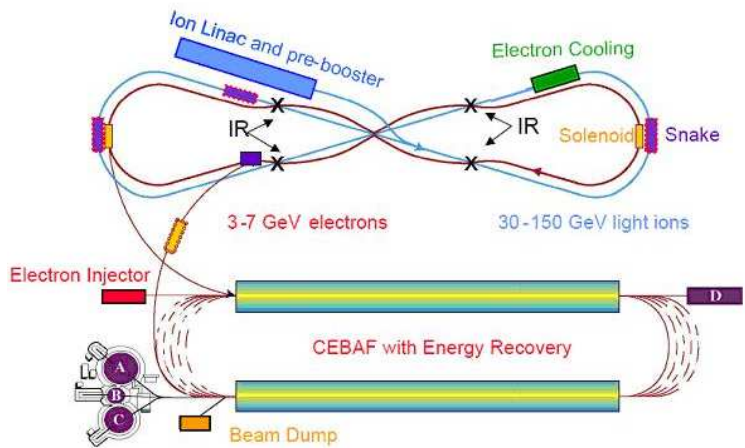}
\caption{Schematic layout of ELIC at Jefferson Laboratory \cite{ELIC web}.}
\end{center}
\end{figure}

\subsection{QCD Explorer (LHeC linac-ring option, CERN)}

QCD Explorer means to construct a moderate energy electron LINAC
(50-100 GeV) tangentially to LHC Ring. This construction will
provide opportunity to utilize highest energy hadron beams for
lepton-hadron collisions. QCD Explorer has two
main goals:

*to get more precision data on PDF's which will be necessary for
adequate interpretation for future LHC data

*to enlighten fundamentals of QCD

For this purpose, the technologies for electron-positron
colliders, which have developed up today can be used or new
technologies can be created.

\subsubsection{CLIC Based}

In this case main problem is occurred by drastically different beam
structure: bunch spacing of LHC is 25 ns comparing with 0.6 ns at CLIC. So that
using CLIC bunches shall have a ratio 1/40. This problem can be
solved by changing beam structure of the LHC or CLIC or both.
Super bunch option was proposed for LHC based ep collider in
\cite{Schulte 2004, Schulte EPAC2004} and L = $10^{31}$ $cm^{-2}s^{-1}$ can be handled by this
way.

\subsubsection{ILC (TESLA) Based}

This option has advantages compare to CLIC; bunch spacing is larger so that it is more suitable for matching with LHC
hadron beam. Estimations show that L = $10^{32}$ $cm^{-2}s^{-1}$ can be handled \cite{Karadeniz 2006, Gladilin 2007}.

\subsubsection{Special e-linac}

In the last two years this option is preferred keeping in mind pulsed mode, CW mode and energy recovery linac. Figure 7 shows different
scenarios for the LHC based linac-ring type ep collider. Luminosity up to $3\times$$10^{32}$ $cm^{-2}s^{-1}$ could be
achieved with pulsed or cw linacs \cite{Zimmermann 2008}. If energy recovery is used, the
luminosity gain depend on recovery efficiency. 90 \% recovery
efficiency results in L = $3\times$$10^{33}$ $cm^{-2}s^{-1}$ (if
recovery reach 98 \% luminosity exceed $10^{34}$
$cm^{-2}s^{-1}$).

\subsection{Energy Frontier (CERN)}

If $E_{c}$ $\geq$ 500 GeV LHC based ep colliders is named as Energy Frontier.
These high energies are suspicion to use energy recovery.
Nevertheless L = $10^{32}$ $cm^{-2}s^{-1}$ seem to be achievable
with pulsed linac \cite{Sultansoy 2005}. It is useful to compare physics search potential of three colliders which can be considered as energy frontiers in foreseen future. Namely,

$\sqrt{s}\ = 14$ TeV pp collider with L = $10^{34}$ $cm^{-2}s^{-1}$ (LHC)

$\sqrt{s}\ = 0.5$ TeV $e^{+}$$e^{-}$ collider with L = $10^{34}$ $cm^{-2}s^{-1}$ (ILC)

$\sqrt{s}\ = 3.7$ TeV ep collider with L = $10^{32}$
$cm^{-2}s^{-1}$ ("ILC" x LHC)

Rough estimations \cite{Sultansoy 1998} show that the total capacity of ep and $\gamma$p options for BSM physics (SUSY, compositness etc) research essentially exceeds that of 0.5 TeV linear collider.

\section{Photon-Hadron Colliders}

In 1980's, the idea of using high energy photon beams, obtained by
Compton backscattering of laser light off a beam of high energy
electrons, was considered for $\gamma$e and $\gamma$$\gamma$
colliders (see review \cite{Telnov 2006} and references therein).
Then the same method was proposed for constructing $\gamma$p
colliders on the base of linac-ring type ep machines in
\cite{Alekhin 1991}. Rough estimations of the main parameters of
$\gamma$p collisions are given in \cite{Sultanov 1989}. The
dependence of these parameters on the distance between
conversion region (CR) and interaction point (IP) was analyzed in \cite{Ciftci
1995}, where some design problems were considered.

It should be noted that $\gamma$p colliders are unique feature of linac-ring ep colliders and could not be constructed on the base of standard ring-ring type ep machines (for arguments see \cite{Sultanov 1989, Ciftci 1995})

This type colliders aren't familiar, so that include many unsolved technical problems. These problems look like to $\gamma$e colliders (see review \cite{Telnov 2006} and references the in). Many studies are completed about $\gamma$e colliders and many solutions are proposed about technical problems, which can be applicable for $\gamma$p colliders, too. The last ones have advantage compare to $\gamma$e colliders. This advantage is the distance between CR and IP [see Figure 8]. In $\gamma$e colliders this distance is very short ($\sim$ mm) but $\gamma$p colliders have 1000 times larger distance ($\sim$ m). Large transverse dimensions of proton bunch ($\sim$ 10 $\mu$m) compare to electron bunch ($\sim$ 3 nm) caused distance difference. High energy $\gamma$ beam contains a lot of residual electrons. These residual electrons have to be separate from $\gamma$ beam. In $\gamma$e colliders this separation is not possible because of short distance between CP and IP. Residual electrons can be separated in $\gamma$p colliders because of longer distance between CP and IP. For this purpose 0.01 Tesla magnetic field is enough to separate the residual electrons.

Different aspects of the THERA based $\gamma$p colliders have been considered in \cite{Ciftci 2001}. In \cite{Aksakal 2007, Aksakal H 2007} Linac*LHC based $\gamma$p colliders have been considered for different linac scenarios.

\begin{figure}
\begin{center}
\includegraphics*[width=10.0cm]{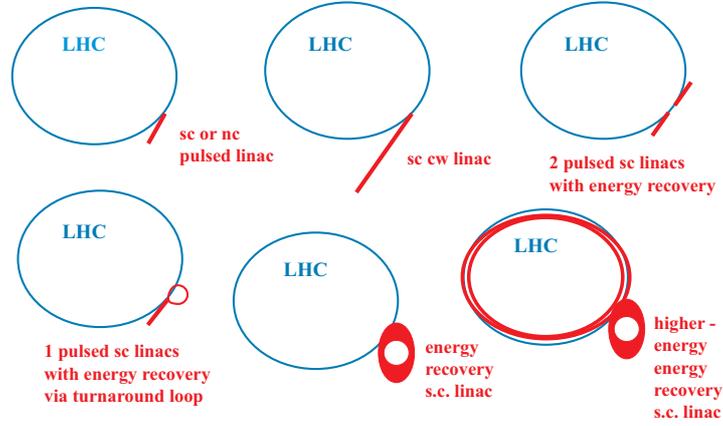}
\caption{Scenarios for the LHC based linac-ring type ep collider \cite{Zimmermann 2008}.}
\end{center}
\end{figure}

\begin{figure}
\begin{center}
\includegraphics*[width=10.0cm]{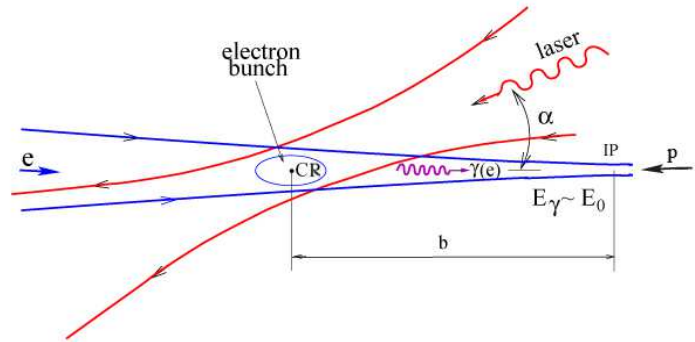}
\caption{Schematic view of $\gamma$p collider.}
\end{center}
\end{figure}

\section{Particle Factories}

As mentioned in Introduction the second purpose of linac-ring type colliders is constructing of high luminosity particle factories, namely, B-factory \cite{Wiesmann 1989,Wiesmann 1990,Cline 1990,Wiesmann 1991}, $\phi$-factory \cite{Cline 1996, Ciftci 1999}, c-$\tau$ factory \cite{Sultansoy 1993, Sultansoy 1995} etc.

Today, linac-ring type B-factory has lost its attractiveness with KEK-B and PEP-B colliders under operation and, especially, Super B proposals. In addition, Super-B factories will copiously produce $\tau$ leptons (the moderate decreasing of the $\tau$ pair production cross section at $\sqrt{s}\approx 10$ GeV is compensated by high luminosity). As a result only a charm factory option of linac-ring type factories still preserves its actuality. In order to search for charm mixing and CP violation by exploiting quantum coherence and to search for rare decays by using a background-free environment, unique opportunities is offered by $\Psi$(3S). Therefore, the center of mass energy is fixed by the mass of $\Psi$(3770) resonance. The CLEO-c worked with L = $10^{32}$ $cm^{-2}s^{-1}$, whereas the BEPC charm factory has design luminosity of $10^{33}$ $cm^{-2}s^{-1}$. Therefore, charm factory with L $>$ $10^{34}$ $cm^{-2}s^{-1}$ will contribute charm physics greatly. It was shown in \cite{Sultansoy PAC2005} that linac-ring option gives opportunity to achieve L = $10^{34}$ $cm^{-2}s^{-1}$. The main restriction on luminosity coming from linac beam power can be relaxed by using of energy recovery linac (ERL). In principle, ERL technology will give opportunity to construct super-charm factory with L well exceeding $10^{35}$ $cm^{-2}s^{-1}$ \cite{Recepoglu 2009}. Linac-ring type charm factory is one of the four main parts of the TAC (Turkic Accelerator Complex) Project \cite{TAC Web}, which is developed since 1997 with the support of Turkish State Planning Organization and planned to be realized before 2020 (see Fig. 9). Recently, a ring-ring tau-charm factory based on the crab waist collision with luminosity of $10^{35}$ $cm^{-2}s^{-1}$ has been proposed at Novosibirsk Budker Institute of Nuclear Physics \cite{Okunev 2008} and high intensity linear $e^{-}e^{+}$ collider for a tau-charm factory with same luminosity is discussed in \cite{Schöning 2007}.

\begin{figure}
\begin{center}
\includegraphics*[width=12.0cm]{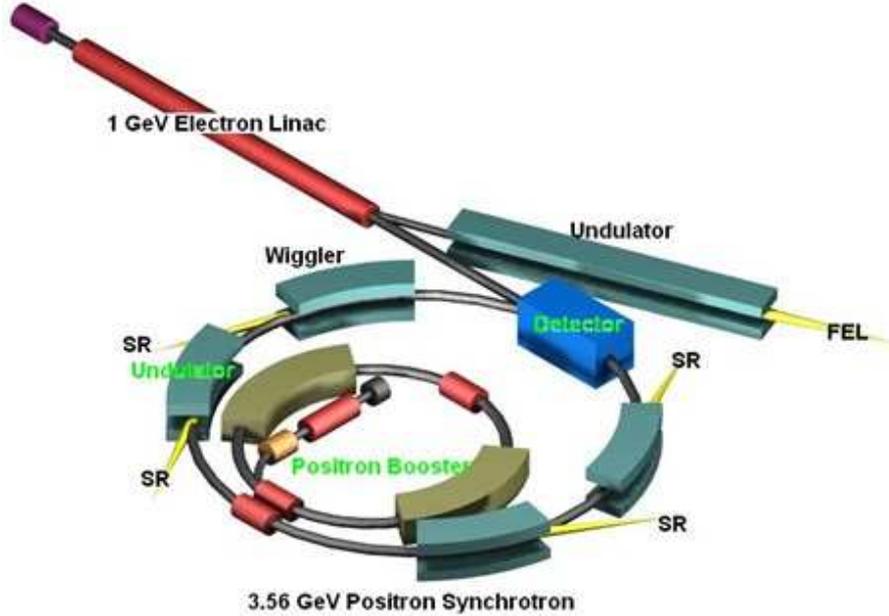}
\caption{Schematic view of the TAC charm factory.}
\end{center}
\end{figure}

Different aspects of linac-ring type lepton-hadron, photon-hadron and electron-positron colliders have been considered in [54-70], too.

\section{Conclusion}

Let us repeat that today linac-ring type colliders present sole
realistic way to TeV scale in lepton-hadron and photon-hadron
collisions. QCD Explorer requires special attention, because it
will be necessary both for exploration of the QCD fundamentals and
adequate interpretation of future LHC data. Especially $\gamma$A
option promises crucial results on strong interactions at all
levels from quarks to nuclei.

Concerning particle factories LR type colliders will provide opportunity to construct super charm
factory with luminosity well exceeding $10^{35}$ $cm^{-2}s^{-1}$.

We appeal to ICFA, ECFA and so on, to organize two worldwide
working groups, one on QCD Explorer and second on Super Charm
Factory. It should be noted that appropriate ERL designs are
crucial for both of them.\\

\textbf{Acknowledgments:} This work is partially supported by
Turkish State Planning Organization under the grand no 2002K120250
and Turkish Atomic Energy Authority. \

\newpage
Appendix 1. Ankara Workshop on Linac-Ring Type ep and $\gamma$p
Colliders\\

1.1. Preface\\

The first International Workshop on Linac-Ring Type ep and gp
Colliders held in Ankara between 9-11 April 1997. The workshop was
organized with the supports from Scientific and Technical Research
Council of Turkey (TUBITAK), Ankara University and Deutsches
Elektronen-Synchrotron DESY Directorate. During the workshop more
than thirty reports have been presented and most of the are
published in this proceedings. The recently proposed lepton-hadron
machines, which open a fourth way to investigate TeV scale
physics, are discussed thoroughly from the machine and physics
aspects.\\

Editors:

S. Atag (Ankara University)

S. Turkoz (Ankara University)

A.U. Yilmazer (Ankara University)\\

International Advisory Committee

M. Atac (FNAL)

S. Ayik (Tennessee University)

E. Boos (Moscow State University)

S. Gershtein (IHEP, Protvino)

L. Okun (ITEP, Moscow)

N.K. Pak (TUBITAK and METU, Ankara)

V. Savrin (Moscow State University)

S. Sultansoy (Ankara University and Azerbaijan Academy of
Sciences)

D. Trines (DESY)

A. Wagner (DESY)

B.H. Wiik (DESY)

C. Yalcin (METU, Ankara)\\

Organizing Committee

S. Atag (Ankara University)

Z.Z. Aydin (Ankara University, Chairman)

A. Celikel (Ankara University)

A.K. Ciftci (Ankara University)

O. Yavas (Ankara University)\\

1.2. Workshop Conclusion and Recommendations

New linac-ring type ep, $\gamma$p and $\mu$p colliders will be
constructed after operation of basic $e^{+}e^{-}$, $\gamma$e,
$\gamma\gamma$, pp and $\mu^{+}\mu^{-}$ colliders. They have
advantages in study of quarks and gluons in the proton because
probing particles (e,$\gamma$, $\mu$) have in general well known
structures. These type of collisions are also optimum for
production and study of some new particles such as leptoquarks.
The expected luminosity of ep and $\gamma$p colliders is lower
than that of above mentioned basic colliders. Nevertheless, there
are a number of physical problems, which can be solved at these
new type colliders. These are

- QCD in the new region of parameters

- Leptoquarks, leptogluons and new contact interactions

- Searching for SUSY and wide spectrum of problems beyond the SM,
etc.

In order to obtain there really new results complementary to those
at basic colliders, the luminosities $L(ep)\geq10^{31}$ and
L($\gamma$p)$\geq10^{30}$ are necessary in units of
$cm^{-2}s^{-1}$ and seem to be sufficient. Higher luminosities
require cooling of the proton beams which needs additional
studies. Concerning the $\mu$p colliders rough estimates give the
luminosity L($\mu$p)$\geq10^{33}$$cm^{-2}s^{-1}$, however this
topic calls for more detailed investigation. \textit{As a result of the
workshop, participants came to the point that it will be useful to
organize two workshops, one on the machine parameters and the
other on the physics research program, during the next year.}

Proceedings of the workshop were published as a special issue of the Turkish Journal of Physics, Vol 22, No 7, 1998.

\end{document}